# Correlations Between Subduction of Linear Oceanic Features and Arc Volcanism Volume Around the Pacific Basin

Claudia Adam[1], Valérie Vidal[2], Pablo Grosse[3], and Mie Ichihara[4]

[1] Geology Department, Kansas State University of Agriculture and Applied Science, Manhattan, KS, USA,
[2] ENSL, CNRS, Laboratoire de Physique, F-69342 Lyon, France
[3] CONICET and Fundación Miguel Lillo, Tucumán, Argentina
[4] Earthquake Research Institute, University of Tokyo, Tokyo, Japan

key points: (1) The correlation between subducting oceanic features and arc volcanism is examined by new analytical and data processing methods. (2) The subduction of oceanic plateaus and most hotspot chains is associated with arc volcanism increase. (3) The subduction of active mid-oceanic ridges is associated with slab windows and arc volcano gaps.

## Abstract

Arc volcanoes, created by magma generated from the dehydration of subducting slabs, show great variability in their sizes and along-arc spatial distributions. In this study, we address a fundamental question, namely, how do subduction zones and volcanic arcs respond to the subduction of "atypical" oceanic lithosphere. We investigate the correlation between the geographical location and volume of arc volcanoes and the subduction of linear oceanic features, including hotspot tracks, oceanic plateaus, volcanic ridges, mid-oceanic ridges, arc volcano chains, and fracture zones, around the Pacific basin. We use multidisciplinary and complementary data sets (topography and bathymetry, seismology and volcano morphometry), and design new analytical and data processing methods. We analyze 35 oceanic linear features. The subduction of three oceanic plateaus and five hotspot chains are clearly associated with volcanism increase, whereas four hotspot chains are related to volcanic gaps. We propose that the patterns of volcanism increase or decrease related to these oceanic features depend on the interplay between chemical (potentially enhancing melting) and thermo-mechanical (inhibiting melting) effects, and/or by the variations of the chemical signatures along hotspot chains. The subduction of volcanic ridges is generally associated with small increases in arc volcanism, which may be accounted for by the fact that these features are highly hydrated and therefore promote melt. The subduction of active mid-oceanic ridges is generally associated with slab windows and arc volcano gaps. No clear inference is found for the subduction of inactive arc ridges.

## 1. Introduction

Subduction zones are one of the main features of plate tectonics, occurring where two plates move toward one another, with one plate diving under the other. The portion of the plate that is being subducted, the slab, pulls the tectonic plate, thus creating the "slab pull", one of the main driving forces of plate tectonics (Forsyth and Uyeda 1975; Conrad and Lithgow-Bertelloni, 2002). A prime characteristic of subduction zones is the development of volcanic arcs along the overriding plate margin, generated by magmatism driven by slab dehydration and subsequent mantle partial melting (e.g., Perfit and Davidson, 2000; LaFemina, 2015).

Large variations along subduction zones in terms of arc volcano density and size have been reported worldwide (e.g., Ben-Avraham and Nur 1980; McGeary et al. 1985; de Bremond d'Ars et al. 1995; White et al. 2006; Grosse et al. 2014). Several hypotheses have been proposed to account for these variations, such as differences in crustal thickness and/or stress fields on the overriding plate, in tectonic setting, in subduction angle and/or rate, and/or in the composition, water content and/or structure of the subducting oceanic plate ( e.g., Cross and Pilger, 1982; Crisp 1984; Dominguez et al., 2000; Katz et al. 2003; Seno and Yamasaki, 2003; Mochizuki et al., 2008; Acocella and Funiciello 2010; Spinelli and Harris, 2011; Völker et al. 2011). Large volcanic gaps, hundreds of kilometers long, are typically found in areas where the angle of the subducting plate is too shallow to accommodate a mantle wedge above (e.g., Gutscher et al., 2000). In particular, volcanic gaps have been related to the subduction of oceanic features, such as plateaus or ridges, which may promote shallow subduction due to their buoyancy and/or modify the temperature and pressure conditions inhibiting the dehydration of hydrous minerals and the partial melting of the mantle (McGeary et al., 1985; Gutscher et al., 2000; van Hunen et al., 2002). Two such gaps are related to the subduction of the Cocos Ridge and the Chile Ridge (McGeary et al., 1985). Along the trench where the Chile Ridge subducts, the ~350-km-long Patagonian Volcanic Gap occurs in between the Southern and Austral volcanic zones of the Andes (e.g., Orihashi et al., 2004; Stern, 2004), whereas the subduction of the Cocos Ridge has been related to a ~175 km gap in the Central American volcanic arc (e.g., Nur and Ben-Avraham 1983; McGeary et al., 1985; Kolarksy et al. 1995). On the other hand, a concentration of volcanoes in the Ecuadorian volcanic arc is found over the subduction of the Carnegie Ridge, which is equal in size and has the same origin as the Cocos Ridge (McGeary et al., 1985; Gutscher et al., 1999a). Several explanations have been proposed for this enhanced volcanism: the recent flattening of the subducted slab and the induced warming of the oceanic crust (Bourdon et al., 2003), a thermal effect due to the plate tearing at both edges of the subducting ridge at shallow angles (Gutscher et al., 1999a), and a strong coupling with the upper plate by the higher ridge elevation (Chiaradia et al., 2009).

Other studies have focused on the impact that subducting oceanic features have on the fluid supply for arc magmatism. For example, Klyuchevskoy, a large volcano in Kamchatka (Russia), is emplaced on the track of the sinking Emperor seamounts, and its large size is thought to be due to an increase in fluid supply to the mantle wedge from these seamounts (Dorendorf et al., 2000). Mahony et al. (2011) also pointed out that the increase in the volume of arc volcanism in Kyushu (Japan) since 6.5 Ma may have occurred under the influence of the fluids from the subducting Kyushu-Palau Ridge. Indeed, the products of the large Aso and Kirishima volcanoes are more strongly affected by fluids than other volcanoes in Kyushu, based on trace element studies (Miyoshi et al., 2008a; 2008b). However, it is yet unclear whether the subduction of the Kyushu-Palau Ridge contributes to the formation of these large volcanoes, or whether it is the cause of the gap between them.

Several regional studies report that the subduction of linear oceanic features are associated with increased arc magmatism (Dorendorf et al., 2000; Mahony et al., 2011; Kimura et al., 2014; Manea et al., 2014; Morell, 2016). Conversely, syntheses investigating the effects of subducting oceanic features on a more global scale generally conclude that the subduction of oceanic features reduces arc magmatism (Cross and Pilger, 1982; Nur and Ben-Avraham, 1983; McGeary et al., 1985; Rosenbaum and Mo, 2011). However, most of these studies do not take into account the subduction zone geometry, such as the direction of the subducting feature relative to the trench, or variations in the slab dip. Furthermore, Rosenbaum and Mo (2011) consider only the arc volcanoes closest to the trench where the subduction of anomalous seafloor is taking place. The authors acknowledge these caveats, and discuss how the dip angle and the orientation of the subducting linear oceanic feature relative to the trench may affect their correlations. They mention, for example, the "orthogonal subduction of the Cocos Ridge." Taking into account the subduction zone geometry to compute the subducted feature prolongation may strongly affect the correlations between arc magmatism and subduction of anomalous seafloor.

This study investigates the correlation between arc volcanoes and the subduction of linear oceanic features of high bathymetric relief around the Pacific Ocean. We consider only subducting features that have a linear, continuous, and clear positive signature in the bathymetric data. These features include hotspot tracks, oceanic plateaus, volcanic ridges, mid-oceanic ridges, arc volcano chains, and a few bathymetrically prominent fracture zones. To constrain these correlations, we use multidisciplinary and complementary data sets (e.g., topography and bathymetry, seismology and volcano volumes), and new analytical and data processing methods. The database developed by Grosse et al. (2014) containing volcano volume estimations has been completed by integrating caldera and underwater volcano volumes. We accurately quantify the topographic anomalies on the oceanic side, and their projection under the arc chain. In particular, we design novel methods to take into account the effects of the subduction zone geometry, the slab dip angle and the direction of these subducting features relative to the trench (section 2). In section 3, we present the results for each volcanic arc segment, with a thorough description and quantification of the subducting linear topographic features within each segment. Finally, section 4 discusses the results considering the origin of the subducting feature (hotspot chain, oceanic plateau, mid-oceanic ridge, fracture zone, etc.) and the possible correlations with increase or decrease in arc volcanism.

## 2. Data and methods

### 2.1 Linear oceanic features

The definition of "linear oceanic feature" varies according to authors. In this study, we use an inclusive definition. We consider "linear oceanic features" to be any kind of linear topography or bathymetric high, such as hotspot tracks, volcanic ridges, oceanic plateaus, inactive or active arc ridges, inactive or active mid-oceanic ridges, and fracture zones. These features are linear and continuous, with a clear trend towards the trench. Our selected features include some oceanic plateaus, such as the Ogasawara Plateau, the Dutton Ridge, and the West Torres Plateau, as their extremity near the trench is clearly elongated along a particular direction. Our synthesis excludes isolated seamounts or discontinuous non-linear blobs or tracks such as the Magellan seamounts. To assess the orientation of the linear oceanic features, we used the bathymetric/topographic grid from Tozer et al. (2019). Their model, SRTM15, has a 15 arc second spatial resolution (approximately 0.5 km).

The 35 linear oceanic features selected for this study are displayed in Figure 1. In Table 1 we report the feature names, the acronyms used in this study, the feature type, the name of the subducting and overriding plates, the geographic coordinates ($X_t$, $Y_t$) of the intersection of the linear feature with the trench, the feature width $w$, the slab dip $\delta$, the trench angle $\tau_E$, the angle $\alpha$ between the linear feature and the perpendicular to the trench (oceanic side) and the angle $\beta$ between the perpendicular to the trench and the projection of the linear feature on the continental side (see Figures 2a and 2b and next section). The width of each linear oceanic feature is assessed from their bathymetric signatures.

### 2.2 Projection

Figure 2a is a sketch of a subducting linear feature. Due to the complexity of plate tectonics, the perpendicular to the trench (dashed line) is not necessarily parallel to the plate kinematic velocity. To account for the subduction zone geometry and to compute the projection of the subducted linear feature on the continental side, a geometrical correction is thus necessary. Let us consider a linear feature oriented with an angle $\alpha$ with respect to the perpendicular to the trench (on the oceanic side). As the plate subducts with a dip angle $\delta$ (Figure 2a), the angle $\beta$ corresponding to the projection of the subducted feature on the surface of the Earth, is obtained through a simple geometrical calculation:

$$\beta = \tan^{-1}\left(\frac{\tan \alpha}{\cos \delta}\right) \quad (1)$$

The angle of the trench, $\tau_E$, is extracted from Bird (2003). The angle $\alpha$ between the perpendicular to the trench and the subducting linear feature (Figure 2a,b) is computed from the bathymetry data (section 2.1 and Table 1, column 11). The slab dip, $\delta$, is computed from Slab2.0 (Hayes et al., 2018), which integrates results from global earthquake monitoring and regional seismotectonic studies to provide three-dimensional geometries of seismically active global subduction zones. The slab morphology varies widely along each subducting zone. To compute the dip corresponding to each of our subducting features, we extract the dip $\delta_{100}$ corresponding to the 100 km depth from Slab2.0, which is the typical depth at which melt occurs (Mibe at al., 1999; Syracuse and Abers, 2006; Syracuse et al. 2010). We then vary a virtual slab dip $\delta_i$ from 0 to 90º, with a step of 1º, and compute the corresponding projection of the subducting feature (continental trajectory, angle $\beta_i$, Figure 2b). For each $\delta_i$, the intersection between the continental trajectory and the 100-km isodepth provides the corresponding value of $\delta_{100}$ at this point. The final continental trajectory (and therefore the final slab dip, $\delta$) is found for $\delta_i = \delta_{100}$ by a minimization algorithm. Slab2.0 does not provide the slab morphology in the southernmost part of South America, where two features subduct. For these features, we use the dip provided by Syracuse and Abers (2006) and Syracuse et al. (2010), who used seismicity to determine the slab characteristics. The dip is reported in Table 1, column 9.

### 2.3 Arc volcanoes database

Volcano selection was based on the Smithsonian Institute Global Volcanism Program (GVP) database of Holocene volcanoes (Siebert et al. 2010). We considered all composite volcanoes (stratovolcanoes, complex volcanoes, shield volcanoes, etc.) and collapse calderas located along the volcanic arcs surrounding the Pacific Ocean and spatially related to one or more of the selected linear oceanic features.

Back-arc and monogenetic-type volcanoes and volcanic fields were excluded from our analysis. A total of 561 volcanoes were considered (Figure 1; Supplementary Table 1), distributed along 14 arc segments: Austral and Southern Andes, Central Andes, Northern Andes, Central America, Cascades, Aleutian-Alaska and Wrangell, Kamchatka-Kuril-Hokkaido, Honshu, Izu-Bonin-Mariana, Kyushu-Ryukyu, New Britain, Solomon, Vanuatu, and Tonga-Kermadec (Figures 3-7).

Volcano volumes were obtained from different sources (see Supplementary Table 1). The main source for composite volcanoes was the database of Grosse et al. (2014). This database gives volume estimates using the SRTM 90 m spatial resolution digital elevation models (DEMs) (Jarvis et al. 2008) and the NETVOLC (used for defining the volcano outlines; Euillades et al., 2013) and MORVOLC (used to compute morphometric parameters; Grosse et al., 2012, 2014) softwares. For composite volcanoes that stand partially or completely underwater (i.e., seamounts, submarine volcanoes, volcano islands), we computed volumes considering manually drawn basal outlines and using MORVOLC and the SRTM30PLUS 1 km spatial resolution DEMs (Becker et al. 2009). For other composite volcanoes not included in the database of Grosse et al. (2014), mostly small edifices, we included volumes given in the literature if available, or calculated rough estimates considering the edifice basal footprint and height measured in GoogleEarth imagery and the equation of a cone. For collapse calderas, we considered volumes given in specific articles, if available, or in the database of Geyer and Marti (2008). For calderas lacking previous volume values, we measured caldera areas with GoogleEarth imagery and roughly estimated their volumes following the best-fit regression equation in a volume vs. area plot of calderas with independently computed values.

It should be noted that the data on volcano numbers and volumes suffers from several uncertainties, namely (1) volcano inclusion is based on the GVP database, which is dynamic and continually updated, with additions/exclusions depending on new data; (2) computed volumes are of the volcano edifices only, i.e., far-reaching deposits, such as tephra fall deposits, are not included in the volume estimates; (3) eroded volumes are likewise not included in the volume estimates. However, given the systematic approach in terms of methodology and DEM sources (Grosse et al., 2012), we consider that our database is robust and comparable at a regional scale. Future efforts can refine the database by incorporating information at a more local scale or on individual volcanoes.

### 2.4 Volcano road and densities

From the arc volcanoes database, we construct a "volcano road", which is a continuous smoothed trend derived from the location of the arc volcanoes (Figure 2c). It is designed with the "median" function available in Matlab (Matlab R2016b, Mathworks®). If the trench has a dominant N-S orientation, such as along South America, the latitude of the volcano road is defined by the latitude of the volcanoes along this segment and the longitude by applying the median filter to longitudes of the arc volcanoes. In the same way, if the trench is dominantly oriented E-W, as along the Aleutian subduction zone, the longitude of the volcano road is defined by the longitude of the volcanoes and the latitude by applying the median function to the vector defined by the volcanoes latitude. The volcano road is then interpolated with a constant step of 10 km (Figure 2c, black line).

We quantitatively take into account the spatial distributions of the arc volcanoes by considering two density metrics: the "volcano density number", $d_n$, and the "volcano volume density", $d_V$. The *volcano density* *number*, $d_n$, is the number of volcanoes per 100 km in a given segment, or window (unit is 1/100 km). To compute it, we translate a circle of diameter *d* (*d*=100 km) along the volcano road (blue circle, Figure 2c). At each step, we count the number of volcanoes within the circle. The *volcano density number* $d_n$ does not consider the volume of each volcano. The volumes are accounted for in the *volcano volume density*, $d_V$, which is the sum of the volcano volumes in a given segment, or window (unit in km$^3$/100 km). It is computed in a similar way, by translating the same circle of diameter *d*=100 km along the volcano road (blue circle, Figure 2c). At each translation step, we sum the volumes of the volcanoes encompassed in the circle and compute $d_V$. The 100 km diameter of the statistic window is arbitrarily fixed. A smaller diameter (e.g., 50 km) provides densities which are too discretized, i.e., the volumes and/or numbers are not averaged enough and display peaks and gaps along the volcano road. On the other hand, a larger circle diameter (e.g., 150 km) smooths too much the volcanoes volume and number densities.

### 3. Results

#### 3.1 Austral and Southern Volcanic Zones of the Andes

Considering both the Austral and Southern volcanic zones of the Andes, 48 volcanoes are found along 2,700 km (Figure 3a), giving a $d_n$ of 1.9 volcanoes/100 km (median 2.0 volcanoes/100 km, Table 2). However, only five volcanoes are located in the southernmost Austral zone of ~1,200 km, giving a $d_n$ of 0.55 volcanoes/100 km, whereas the Southern zone has 43 volcanoes and a $d_n$ between 1 and 7 volcanoes/100 km (average of 2.9 volcanoes/100 km). Volcano volumes vary between ~1 and 1,000 km$^3$, with a mean of 82 km$^3$ and a median of 18 km$^3$. Five volcanoes have volumes > 100 km$^3$: Cerro Hudson, Puyehue-Cordón Caulle and the calderas Laguna del Maule, Calabozos and Atuel. The mean $d_V$ is 9.4 km$^3$/100 km in the Austral zone and 265 km$^3$/100 km in the Southern zone, where it varies between 10 and 1,500 km$^3$/100 km. $d_V$ for the whole region is 160 km$^3$/100 km (median 42 km$^3$/100 km, Table 2) and shows a general increase northwards, as previously identified by Volker et al. (2011).

The Phoenix Ridge is an inactive spreading center (Livermore et al., 2000; 2004; Rosenbaum and Mo, 2011). It intersects the trench SE of the Austral volcanic zone where no volcanoes are found (Figures 3a,b); hence, no volcanism seems to be related to this linear anomaly.

The Chile Ridge is an active mid-oceanic ridge (Cande et al., 1987; Lagabrielle et al., 1994, 2000; Tebbens et al., 1997, Russo et al., 2010; Bourgois et al., 2016). Its subduction coincides with the Patagonian Volcanic Gap between the Austral and Southern volcanic zones at 46-49°S, although its wide swath includes Cerro Hudson, Lautaro and Viedma volcanoes (Figures 3a,c). Aguilera is located outside of the projection swath, close to its southern margin (Figure 3c). Subduction of the Chile Ridge has been considered responsible for this gap in arc volcanism and for back-arc alkaline basaltic volcanism (e.g., Gorring and Kay, 2001; Stern, 2004). The projection of this linear trend actually coincides with the location of the supposed Arenales volcano (Lliboutry, 1999) (Figure 3c), although its origin has not been confirmed (Stern et al., 2016). Therefore, we do not include Arenales for the densities computation. We find an increase on the sides and a decrease in the middle in both $d_n$ and $d_V$ in the region corresponding to the projection of the Chile Ridge (Figure 3e).

The Juan Fernandez Ridge is a hotspot chain (Juan Fernandez hotspot; Bello-González et al. 2018). It intersects north of the Southern volcanic zone, where no volcanoes are located ($d_n$ and $d_V$ are null) (Figures 3a, d). Subduction of this ridge has been related to the decrease in the subduction angle of the Pampean flat slab (27-33°S) and the volcanic gap between the Southern Central volcanic zones of the Andes (e.g., Pilger, 1981; Kay and Mpodozis, 2002; Pardo et al., 2002; Stern, 2004; Anderson et al., 2007; Bello-González et al., 2018).

### 3.2 Central Volcanic Zone of the Andes (CVZ)

This segment of 1,270 km contains 52 volcanoes with $d_n$ between 0 and 8 volcanoes/100 km and an average of 2.9 volcanoes/100 km (median 3.0 volcanoes/100 km, Figures 3f-k and Table 2). Volcano volumes vary between ~1 and 200 km$^3$ with mean and median averages of 36 km$^3$ and 19 km$^3$, respectively. Six volcanoes have volumes > 100 km$^3$. $d_V$ varies between 0 and 380 km$^3$/100 km with an average of 105 km$^3$/100 km (median 72 km$^3$/100 km, Table 2). Two gaps in both volcano number and volume are found at ~26ºS and ~20ºS (Figures 3f, k).

The Copiapó Ridge, also known as the Easter Hot Line, is a hotspot chain created by the Caldera hotspot (Bello-González et al., 2018). Its projection coincides with the southernmost segment of the Central Volcanic Zone that shows peaks in both $d_n$ and $d_V$ (Figures 3g, k), as has been previously documented (e.g., González-Ferrán et al. 1985; Grosse et al. 2018). In this region between latitudes ~26.5 − 27.3°S, nine small to large volcanoes adding up 385 km$^3$ are found along only ~70 km (Figure 3g), and several more recent volcanoes have been inferred (e.g., Grosse et al. 2018). South of this region, volcanism is absent whereas to the north volcanism decreases towards the volcanic gap at ~26ºS. Subduction of the Copiapó Ridge has been linked to a regional tectono-volcanic discontinuity at ~27°S, abundant ENE-WSW structures (parallel to the projection of the linear feature) and an expansion of volcanism towards the back-arc (e.g., Bonatti et al. 1977; González-Ferrán et al. 1985; Álvarez et al. 2015). Furthermore, Álvarez et al. (2015) propose that the Copiapó Ridge controlled the northern edge of the Pampean flat slab, similar to the control that the Juan Fernández Ridge exerts at the southern edge of the flat slab.

The San Félix volcanic chain, also called Taltal Ridge is created by the San Félix hotspot (Bello-González et al., 2018). Its central projection intersects Llullaillaco volcano and runs in between a cluster of smaller volcanoes to the south (Cerro Escorial, Lastarria, Cordón del Azufre, Cerro Bayo) and some large volcanoes to the north (Socompa, Pular, Aracar) (Figures 3f, h). The swath of the linear feature coincides with a decrease in $d_n$ in the middle of the box, whereas it increases on both sides; $d_V$ increases on the northern side of the box, around latitude 24°S (Figure 3k).

The Iquique Ridge is a hotspot trail created by the Foundation hotspot (Bello-González et al., 2018). The projection of this feature spans a ~170 km region between ~19.2-20.7°S without recent arc volcanism, known as the Pica Gap (Figures 3f, i) (Wörner et al., 1992). Wörner et al. (2000) have suggested a link between this gap in volcanism and the subduction of the Iquique Ridge. However, the swath of the linear feature also includes six volcanoes north of this gap (Figure 3i). $d_n$ (1 to 4 volcanoes/100 km) and $d_V$ (10-46 km$^3$/100 km) in this region are below the averages of the Central Volcanic Zone (Figure 3k).

The Nazca Ridge is a hotspot trail created by the Salas y Gómez hotspot (Bello-González et al., 2018). The Mendana Fracture Zone (Huchon and Bourgois, 1990), located north, is not represented in Figure 3 but can be seen in Figure 1. The projections of both features cross well north of the Central Volcanic Zone, within the Peruvian flat slab segment, where no volcanism is located (Figures 3f, j). The low-angle subduction in this region has been linked to a relatively buoyant Nazca Ridge within the subducting slab (e.g., Pilger 1981; Gutscher et al., 1999b, 2000; Rosenbaum et al., 2005; Bishop et al. 2017).

### 3.3 Northern Volcanic Zone of the Andes

This segment of 860 km contains 30 volcanoes. $d_n$ varies between 1 and 8 volcanoes/100 km, with an average of 2.9 volcanoes/100 km (median 2.0 volcanoes/100 km, Table 2). Volcano volumes vary between 2 and 670 km$^3$, with mean and median averages of 74 km$^3$ and 46 km$^3$, respectively. Six composite volcanoes have volumes above 100 km$^3$, and one caldera, Chacana, has the largest volume of 670 km$^3$. Average $d_V$ is 209 km$^3$/100 km (median 121 km$^3$/100 km); in most of the region, $d_V$ is between $50-230$ km$^3$/100 km, with a peak above 1,200 km$^3$/100 km at ~0.3°S (where Chacana and Guagua Pichincha are located).

The Sarmiento and Alvarado ridges are fossil volcanic ridges created by fissure eruptions (Lonsdale, 2005; Yepes et al., 2016). Northwards are a scarp south of the Grijalva Fracture Zone (FZ) and the Grijalva Fracture Zone (Figure 4a), whose origin is debated. According to Gutscher et al. (1999a), it represents the southern half of the scar where the Nazca Plate tore off. For Yepes et al. (2016), the topographic step across this fault zone is due to the density contrast between the younger Nazca crust to the north and the older Farallon crust to the south. In any case, this linear feature is not associated with OIB (Ocean Island Basalts) volcanism, but rather with a topographic step.

These four linear features cross the southern part of the Northern Volcanic Zone where three volcanoes (Sangay, Chimborazo, and Tungurahua) are located rather isolated from the rest of the volcanoes to the north (Figures 4a, g). The Sarmiento Ridge projects south of the arc segment where no volcanism is present (Figure 4b); The Alvarado Ridge projects north of Sangay and includes Tungurahua (Figure 4c); the scarp south of the Grijalva Fracture Zone projects on Chimborazo (Figure 4d); and the Grijalva Fracture Zone projects in between these three volcanoes and the cluster of volcanoes to the north, with its swath including a couple of these volcanoes (Cotopaxi and Antisana, Figure 4e). Taken together, the southern three linear features can be related to the three relatively isolated volcanoes at the segment's southern margin, although these volcanoes could also be related to the Carnegie Ridge (see below). The Grijalva Fracture Zone separates an older portion of the Nazca plate to the south from a younger portion of the Nazca plate to the north (e.g., Yepes et al. 2016; Ancellin et al. 2017). Ancellin et al. (2017) propose that the older and probably colder subducting plate south of the Grijalva Fracture Zone may dehydrate and/or melt to a lesser extent. This may be related to the relative scarcity of volcanoes in the southernmost part of the Northern Volcanic Zone (Figure 4a). A peak in $d_n$ and $d_V$ can be observed where the scarp south of the Grijalva Fracture Zone and the Grijalva Fracture Zone overlap, at around 1.1°S (Figure 4g).

The wide Carnegie Ridge subducts between latitudes 2.5°S and 0.5°N (Gutscher et al. 1999a). Rosenbaum and Mo (2011) and Yepes et al. (2016) consider the Carnegie and Cocos ridges as aseismic ridges, whereas other studies provide more information about their origin. For example, the Carnegie and Cocos ridges are hotspot tracks created respectively by the eastward motion of the Nazca Plate and the northeastward motion of the Cocos Plate over the Galapagos plume (Gutscher et al., 1999a; Meschede and Barckhausen, 2001; Lonsdale, 2005; Morell, 2015). The central trace of the Carnegie Ridge projects in between the three southern volcanoes and a cluster of volcanoes to the north (Figure 4f), but its wide swath includes both the volcanoes to the south and north, in total 16 volcanoes, giving higher than average $d_n$ of 3.8 volcanoes/100 km (median 3.0 volcanoes/100 km) and $d_V$ of 388 km$^3$/100 km (median of 175 km$^3$/100 km), compared to the whole Northern Volcanic Zone (Figure 4g and Table 2). The northern part of the swath contains a large peak in both $d_n$ and $d_V$ (Figure 4g). Several studies have investigated the influence of the subduction of the Carnegie Ridge on the geochemical signature of the Ecuadorian arc (e.g., Bourdon et al., 2003; Samaniego et al. 2005; Hidalgo et al. 2007, 2012; Chiaradia et al. 2009, 2020). It has been proposed that subduction of this ridge produces a shallowing of the subduction angle, induced by young and buoyant oceanic lithosphere, which in turn favors a high geothermal gradient that prompts slab partial melting (e.g., Gutscher et al. 1999a; Bourdon et al. 2003). The gaps in volcanism in the northern part of the Northern Volcanic Zone do not seem related to any linear oceanic feature.

The Cocos Nazca Ridge is an active mid-oceanic ridge between the Northern Volcanic Zone and Central America (Figure 4h) (Johnston and Thorkelson, 1997; Abratis and Wörner, 2001; Wang et al., 2020). It is associated with a gap in volcanism.

### 3.4 Central America

This segment runs for almost 2,000 km and contains 52 volcanoes (Figure 4i), with overall average $d_n$ and $d_V$ of 2.8 volcanoes/100 km and 123 km$^3$/100 km (median values of 3.0 volcanoes/100 km and 83 km$^3$/100 km, Table 2). Volcano volumes range between < 1 to 300 km$^3$, with mean and median averages of 44 km$^3$ and 17 km$^3$, respectively. The southernmost section of over 500 km contains only two large isolated volcanoes, El Valle and Barú (Panama) (Figures 4i, j). Likewise, small El Chichón is the only volcano in the northernmost 250 km section (Figures 4i, k). The main section of the arc has $d_n$ between 2 and 8 volcanoes/100 km (Figure 4l). Volume densities $d_V$ are mostly between 25 and 220 km$^3$/100 km, with the exception of a peak of 1,000 km$^3$/100 km in the southern portion of the main section where the five largest composite volcanoes of Central America (each with volumes > 95 km$^3$) are located along the Central Mountains of Costa Rica (Figures 4i, l).

As described above, the Cocos Ridge is created by the interaction between a mid-oceanic ridge and the Galapagos plume (Gutscher et al., 1999a). The Cocos Ridge is a very wide feature that crosses the southernmost section of the arc (Figures 4 i, j). Its central trace projects on Barú volcano, and falls just short of covering the peak of large volcanoes of Costa Rica to the north (Figures 4 j, l). El Valle volcano is located just SE of the feature swath (Figure 4j). Several studies suggest that the subduction of the Cocos Ridge inhibited magmatism and is the cause of the paucity of volcanism in this area (e.g., Nur and Ben-Avraham, 1983; McGeary et al., 1985; Kolarksy et al., 1995). On the other hand, Hidalgo and Rooney (2010) propose a link between the subduction of the Cocos Ridge and the adakitic signature of lavas erupted at Barú.

An unnamed transform fault corresponds to the last surface expression of the mid-oceanic ridge between the Cocos and Nazca plates (Figures 4h,i). It subducts under the Caribbean plate (Bird, 2003; Manea et al., 2014) and projects in between Barú and the large volcanoes of Costa Rica (Figure 4j).

The Seamount Province (Werner and Hoernle, 2003; Harpp et al., 2005) projects through this gap (although its swath touches the two southernmost Costa Rica volcanoes, Irazú and Turrialba) and within the swath of the Cocos Ridge projection (Figure 4j). Subduction of the Fischer Seamount chain north of the Seamount Province could be related to the volumetric peak in volcanism given by the five southernmost volcanoes of Costa Rica (e.g., von Huene et al., 2000). However, we did not include it in our analysis because it is not a continuous linear feature. Benjamin et al. (2007) relate the unusual composition of Irazú (enrichment in incompatible trace elements, high volatile contents) with subduction of seamount chains.

The Tehuantepec Ridge is a fracture zone (Manea et al., 2005; Rosenbaum and Mo, 2011) that projects well north of the main arc section, but includes El Chichón volcano on its northern swath margin (Figure 4k). Manea and Manea (2008) and Manea et al. (2014) propose that water-rich magmas from El Chichón reflect locally anomalous fluxes of slab-derived fluids associated with the subduction of Tehuantepec Ridge.
The transform fault, the Tehuantepec Ridge and the Cocos Ridge are associated with a decrease in both $d_n$ and $d_V$, although two local peaks exist in $d_n$ and $d_V$ in the swath of the Cocos Ridge (Figure 4l). Although the Seamount Province seems associated with a local minimum, the median value of $d_V$ in the swath is (very) slightly larger that the arc median (see Table 2). The Tehuantepec Ridge is associated with a small peak in $d_n$ and $d_V$, although the values of $d_n$ and $d_V$ in the corresponding box are smaller than the average values for the volcanic arc (Table 2).

### 3.5. Cascades Volcanic Arc

The Cascades Volcanic Arc spans 1,180 km and contains 19 volcanoes (Figure 5a). $d_n$ and $d_V$ are relatively small (Figure 5b), with mean and median values of 1.4 and 1 volcanoes/100 km ($d_n$), and 60 and 47 km$^3$/100 km ($d_V$). No linear feature subducts under this arc.

The Juan de Fuca Ridge is an active mid-oceanic ridge that subducts north of the Cascades, between the Cascades and the Wrangell volcanic arcs (Cole and Stewart, 2009; Thorkelson et al., 2011). It is associated with a gap in volcanism (Figure 5a).

### 3.6. Aleutian-Alaska and Wrangell Volcanic Arcs

The Aleutian-Alaska and Wrangell arcs form a vast segment of over 3,000 km with 71 volcanoes (Figure 5c). $d_n$ varies from 0 to 11 volcanoes/100 km with an average of 2.3/100 km (median 2.0/100 km, Table 2). Volcano volumes vary between < 1 to 1,100 km$^3$, with mean and median averages of 71 km$^3$ and 19 km$^3$, respectively. Ten volcanoes, including one caldera, have volumes above 100 km$^3$. $d_V$ is mostly between 10 and 500 km$^3$/100 km, with an average of 145 km$^3$/100 km (median 64 km$^3$/100 km, Table 2). The most notable exception is at the eastern margin, in the Wrangell Arc, in the Alaska interior, where three very large volcanoes (Churchill, Wrangell and Sanford) combine to give volume densities above 1,000 km$^3$/100 km (Figure 5f).

The Yakutat microplate or block, identified by some authors as an oceanic plateau (Eberhart-Phillips et al., 2006; Gulick et al., 2007), subducts under the Wrangell Arc. Although we have not included it in our analysis because it is not a linear feature, Koons et al. (2010) indicate that the Yakutat block is located between longitudes 215 and 225°E under our volcano road (Figure 5f). There is a small increase in $d_n$ related to this feature and a very large increase in $d_V$, as some of the world's largest arc volcanoes are found above this plate (Churchill, Wrangell and Sanford; Figures 5c, f).

The Kodiak-Bowie Chain and the Cobb Seamounts are hotspot chains that intercept the Alaska subduction zone (Figure 5c), where the oldest extremities of the chains (respectively 23.8 and 29.2 Ma for Kodiak-Bowie and Cobb) are subducting (Turner et al., 1973, 1980; Dalrymple et al., 1987; Desonie and Duncan, 1990; Clouard and Bonneville, 2005). The projection of the Kodiak-Bowie Chain intersects a section of the arc in the Alaska Peninsula where several small to intermediate volcanoes are found (Peulik, Griggs, Snowy Mountain, Figure 5d). Its subduction is associated with an increase in $d_n$ and a decrease in $d_V$, although both display local peaks in the projected tracks (Figure 5f).

The projection of the Cobb Seamounts coincides with the Veniaminof and Kupreanof volcanoes (Figure 5e). Veniaminof has a volume of 344 km$^3$, making it by far the largest volcano of the segment (excluding the three volcanoes of the Wrangell Arc) and produces a large peak in $d_V$, and a small peak in $d_n$ (Figure 5f).

### 3.7. Kamchatka-Kuril-Hokkaido Volcanic Arc

This 2,200 km segment, also called Kamchatka or Kamchatka-Kuril, contains 105 volcanoes giving one of the highest average $d_n$ of 4.4 volcanoes/100 km (median 4.0/100 km, Figures 5g-j and Table 2). $d_n$ varies mostly between 1 and 10 volcanoes/100 km (Figure 5j). $d_V$ vary between < 1 km$^3$ to 300 km$^3$, with mean and median averages of 41 km$^3$ and 23 km$^3$, respectively (Figure 5j). 13 volcanoes, including three calderas, have volumes above 100 km$^3$. $d_V$ varies between 0 and 500 km$^3$/100 km, with a mean and a median of 179 and 149 km$^3$/100 km, respectively (Figure 5j).

The Hawaii-Emperor Chain is one of the largest and longest-lived hotspot chains on Earth (Morgan, 1971; Sleep, 1990; King and Adam, 2014). The oldest (75.8 Ma) and northwesternmost end of the Emperor Seamounts is subducting under the northernmost part of the Kuril-Kamchatka trench, at longitude 168°W (Clague and Dalrymple, 1989; Duncan and Keller, 2004; Clouard and Bonneville, 2005). Its projection correlates with 15 volcanoes, including the Klyuchevskoy volcanic group (Figure 5h), that produces a peak in both $d_n$ and $d_n$ (Figure 5j). This linear feature also seems to correlate with an across-arc shift in volcanism, since the Klyuchevskoy group is displaced towards the interior (i.e., away from the trench), and also with a large peninsula jutting to the sea. Dorendorf et al. (2000) propose that subduction of the hydrothermally altered Emperor Seamounts is responsible for the fluid-rich compositions of the Klyuchevskoy volcano products. Recently, Nishizawa et al. (2017) found remarkable chemical variations at monogenic volcanic cones at the forearc of this region and attributed it to the Emperor Seamounts subduction.

The next linear feature towards the south corresponds to the Morozko Seamounts (Figure 5g) (El Hariri et al., 2013). We classify it as a volcanic ridge, although its origin is still debated. Some studies propose that it has a similar origin as the Petit-Spot seamounts, located southwest of the Morozko Seamounts. For Petit-Spot seamounts, the extension stress field created in the lithosphere by the bending of the subducting plate facilitates the ascent of magma (Hirano et al., 2001, 2006; Hirano, 2011). Khomich et al. (2019) proposed that subduction of such Petit-Spot volcanoes may influence the formation of ore-magmatic systems in South Kuril. The trace of the Morozko Seamounts encompasses 11 volcanoes, including the two most voluminous volcanoes of the region on either of its sides (Chirinkotan and Alaid) (Figure 5i). The Morozko Seamounts subduction is associated with a local peak in $d_n$, although $d_n$ is smaller than the average or median value along the segment (Table 2). $d_V$ shows a significant increase at the box sides, and a decrease in the middle (Figure 5j). For this feature, $d_V$ is larger than the median $d_V$ along the volcanic arc (Table 2).

### 3.8 Honshu Volcanic Arc

This 650 km segment contains 33 volcanoes (Figure 6a). $d_n$ varies mostly between 1 and 6 volcanoes/100 km (Figure 6b), giving an average density of 2.8 volcanoes/100 km (median 3.0 volcanoes/100 km, Table 2). Volcano volumes vary between < 1 km$^3$ to 180 km$^3$, with mean and median averages of 41 km$^3$ and 26 km$^3$, respectively. Four volcanoes have volumes above 100 km$^3$. $d_V$ varies between 10 and 350 km$^3$/100 km, with an average of 96 km$^3$/100 km (median 80 km$^3$/100 km, Figure 6b and Table 2).

The Izu-Bonin-Mariana Arc (Figure 6c) is an active oceanic island arc developed on the Philippine Sea Plate (e.g., Stern et al., 2003; Takahashi et al., 2007). An early arc split off to form the Izu-Bonin-Mariana Arc to the east and the Kyushu-Palau Ridge to the west, during the formation of now inactive Shikoku and Parece Vela Basins (Stern et al., 2003; Okino et al., 2015). The Izu-Bonin-Mariana Arc has been colliding with central Japan during the last 15 Ma (Takahashi and Saito, 1997). Two large volcanoes, Mt. Fuji and Hakone, are located at this collision. Geochemical studies reveal that magmatism at Mt. Fuji is driven by the fluid released from the Pacific Plate (e.g., Ichihara et al., 2017 and references therein). Earlier studies assumed that Mt. Fuji is large because the subducted Philippine Sea Plate is teared beneath it (Takahashi, 2000). More recent seismological studies have resolved the structure beneath and beyond Mt. Fuji and show a low-velocity zone (Kinoshita et al., 2015) and a cluster of seismicity (Kashiwagi et al., 2020) in the subducted Philippine Sea Plate, right beneath Mt. Fuji. They may be a remnant of a magmatic system of what was an arc volcano of the Izu-Bonin-Mariana Arc, which supplies excess magma to Mt. Fuji (Ichihara et al., 2017). The projection of the Izu-Bonin-Mariana Arc coincides with the off-arc Mt. Fuji and Hakone volcanoes (Figure 6a). Its swath is very wide and approximately coincides with a broad peak in $d_n$ and $d_V$ (Figure 6b).

### 3.9 Izu-Bonin-Mariana Arc

This 2,800 km segment contains 51 volcanoes (Figures 6c-f). $d_n$ are rather low, ranging from 0 to 6 volcanoes/100 km and with an average of 1.8 volcanoes/100 km (median 2.0 volcanoes/100 km, Table 2). Volcano volumes vary between 1 km$^3$ to 650 km$^3$, with high mean and median averages of 121 km$^3$ and 90 km$^3$, respectively. Volumes tend to be greater in the southern section (Marianas), where several volcanoes have volumes above 100 km$^3$ (Figure 6c). The average $d_V$ for this segment is 233 km$^3$/100 km (median 131 km$^3$/100 km, Table 2).

The Ogasawara Plateau formed by late Cretaceous igneous activity within the Pacific Plate (Okamura et al., 1992). It formed 110-80 Ma ago, but a new volcanic stage (~55 Ma) overprinted the older volcanism in the area (Hirano et al., 2021). The geochemical signature of the lavas along this arc reflects the subduction of OIB (Ocean Island Basalt), including HIMU (high U/Pb) material (Plank and Langmuir, 1998; Sano et al., 2016). The projection of the Ogasawara Plateau crosses the arc close to the rather large Kaitoku seamount (140 km$^3$) and its swath includes the very large Doyo seamount (340 km$^3$) to the north (Figures 6c, d). $d_n$ is around average in this section, whereas a peak in $d_V$ on the northern part of the swath is caused by the Doyo seamount and neighboring Nishino-shima (Figure 6g).

The Dutton Ridge is an oceanic plateau overprinted with guyots (Smoot, 1983) (Figures 6c, e). Its projection crosses the northern segment of the Marianas Arc. Its large swath includes two peaks in both $d_n$ and $d_V$, and the projection center-line coincides with a gap between these two peaks (Figure 6g). Two small volcanoes west of the main volcano road are encompassed in the Dutton Ridge swath (Figure 6e).

The Caroline Ridge (or Caroline Islands) is a hotspot chain (Ito and van Keken, 2007; Clouard and Bonneville, 2005; King and Adam, 2014). The age of the oldest volcanoes close to the trench where the Pacific Plate is subducting is 13.9 Ma (Ito and van Keken, 2007). The Caroline Ridge convergence is thought to have resisted subduction and produced a collision (e.g., McGeary et al., 1985). No volcanism is present in the region (Figures 6c, f).

### 3.10 Kyushu and Ryukyu Islands

This segment contains 18 volcanoes along 1,180 km (Figure 6h), and has an average and median $d_n$ of 1.5 volcanoes/100 km and 1.0 volcanoes/100 km, respectively. Volcano volumes vary between 1 and 410 km$^3$, with mean and median averages of 81 km$^3$ and 44 km$^3$, respectively. The two largest volcanoes are the Aira caldera (Sakurajima) and Aso, each with volumes of ca. 400 km$^3$. $d_V$ is 128 km$^3$/100 km on average, with a median value of 45 km$^3$/100 km.

The Kyushu-Palau Ridge is an ancient arc that was active at 25-48 Ma, and most of the volcanic rocks currently exposed on the seafloor were dated to 25-28 Ma (Ishizuka et al., 2011). Its projection crosses the northern part of the segment, including Kuju and Aso volcanoes, and passes north of a volcanic gap between Aso and the southern volcanoes (Figure 6h, i). It is associated with an increase in $d_n$ and $d_V$ (Abe et al., 2013; Hata et al., 2014) (Figure 6m).

The Oki Daito Ridge, Daito Ridge, and Amami Plateau are inactive island arcs (Figure 6h). Their ages are debated. They are Cretaceous (145-66 Ma) features according to Tokuyama (1995), Hickey-Vargas (2005) and Ishizuka et al. (2011), whereas more recent studies show that Jurassic to Cretaceous (201.3-66 Ma) arc crust is overprinted by Eocene (56 Ma-33.9 Ma) volcanism (Okino, 2015). The Amami Plateau and Daito Ridge were separated from a single volcanic arc by a back-arc rifting and formation of the North Daito Basin from 50-40 Ma to 30-33 Ma (Okino, 2015; Deschamps and Lallemand, 2002). On the other hand, the Oki Daito Ridge was formed separately and approached to Daito Ridge from the far south (Higuchi et al., 2015).

Okamura et al. (2017) investigated the effect of these ridge subductions to the forearc structure, and show that the Daito Ridge, the Amami Plateau and the Kyushu-Palau Ridge migrated North and West in the last 23 Ma, but did not discuss their effects on volcanism. The bathymetric signature of the Oki Daito Ridge stops almost 200 km before it reaches the trench. As we are not sure this inactive arc has already subducted, we do not include it in our analysis. On the other hand, the bathymetry data clearly shows a thin plateau situated south of the Amami Plateau. We include it in our synthesis, and call it South Amami Ridge. The swath of the the Amami Plateau just touches Iwo-Tori-shima (Figure 6j), which is the southernmost and quite isolated volcano of the segment. The projection of the South Amami Ridge crosses between Yokoate-jima and Iwo-Tori-shima and does not include any volcano (Figure 6k), leading to a gap in $d_n$ and $d_V$ (Figure 6m). The Daito Ridge crosses the segment well south of the southernmost volcano and hence is not associated to volcanism (Figure 6l).

### 3.11 New Britain Arc

The New Britain volcanic arc has a length of 1,070 km and contains 29 volcanoes. The average $d_n$ is 2.4 volcanoes/100 km (median 2.0 volcanoes/100 km, Table 2). Volcano volumes vary between <1 and 1,000 km$^3$, with mean and median averages of 105 km$^3$ and 16 km$^3$, respectively. $d_V$ is 264 km$^3$/100 km on average, with a median value of 92 km$^3$/100 km (Table 2). No linear feature subducts under this arc.

### 3.12 Solomon Arc

This segment spans 1,100 km and contains 11 volcanoes (Figure 7a), giving an average $d_n$ of 1.1 volcanoes/100 km (median 1.0 volcanoes/100 km). Volcano volumes vary between 2 and 830 km$^3$, with mean and median averages of 159 km$^3$ and 21 km$^3$, respectively. Volcano volume densities range from 0 to 830 km$^3$/100 km, with an average of 185 km$^3$/100 km and a much lower median value of 29 km$^3$/100 km (Table 2).

The Woodlark Spreading Center is an active mid-ocean ridge subducting northeastward beneath the Solomon Arc (König et al., 2007; Chadwick et al., 2009; Holm et al., 2016; Wang et al., 2020) (Figure 7a). Its projection encompasses three volcanoes: Savo, Kavachi, and an unnamed volcano (Figure 7a). Another three volcanoes, Simbo, Kana Keoki and Coleman Seamount, are emplaced on the subducting plate (Figure 7a), and are therefore not considered for density calculations (see discussion in section 4.3). The Woodlark Spreading Center subduction is associated with two peaks in $d_n$, created by the Savo, Kavachi, and the unnamed volcano. There is no noticeable increase in $d_V$, as the previously discussed volcanoes have a relatively small volume. $d_V$ is almost null along the Woodlark Spreading Center swath (Figure 7b).

### 3.13 Vanuatu Arc

This segment contains 18 volcanoes along 1,530 km (Figures 7c-f), giving a low $d_n$ of 1.1 volcanoes/100 km on average (median 1.0 volcanoes/100 km, Table 2). Mean and median average volumes are 372 km$^3$ and 71 km$^3$, respectively, and several volcanoes have volumes > 100 km$^3$, including three with volumes > 1,000 km$^3$ (Gaua, Aoba and Aneityum). $d_V$ is 422 km$^3$/100 km on average (Table 2). The median volume density is much lower, 5.6 km$^3$/100 km, due to the presence of two large peaks of 1,780 and 2,300 km$^3$/100 km in the northern region (Figure 7f).

The West Torres Plateau is an oceanic plateau formed by a mantle plume during either the Oligocene (33.9-23 Ma) (Yan and Kroenke, 1993), or prior to the Eocene and collided with the subduction zone in the late Eocene (Wells, 1989; Meffre and Crawford, 2001). The subduction of the West Torres Plateau started at around 0.7 Ma (Meffre and Crawford, 2001). Its projection coincides with the largest $d_V$ peak of the segment, related to four volcanoes, most notably Gaua (Figures 7d, f).

The D'Entrecasteaux Ridge consists of two parallel ridges which are inactive volcanic arcs (Baillard et al., 2018, Beier et al., 2018; Meffre and Crawford, 2001). The southern D'Entrecasteaux Ridge is composed of discrete extinct volcanoes with primitive island arc lavas, whose magmatism stopped at 36 Ma. The northern D'Entrecasteaux Ridge has a smoother surface and consists of lavas resembling those in the modern Mariana forearc. The D'Entrecasteaux Ridge began colliding with the Vanuatu Arc at around 2 Ma and the collision is migrating northwards with time at 20-40 mm/yr (Meffre and Crawford, 2001).The projection of this feature coincides with Aoba volcano, the largest volcano of the segment and the cause of the second peak in $d_V$ (Figures 7e, f). Aoba is rather isolated, located in a window of 180 km without any other volcano, and hence no peak in $d_n$ is associated with this feature (Figure 7f). Our method assumes that the projection of the linear features are straight, which can introduce an error if the feature actually has a curved projection, as has been postulated for the D'Entrecasteaux Ridge (Baillard et al., 2018). The curved projection proposed by Baillard et al. (2018) is based on a tectonic model, assuming a temporal migration of this feature. When the authors use only seismic data to track the subducted features, the projections are straight lines, in agreement with the ones proposed here (see section 3.14 for example).

### 3.14 Tonga and Kermadec Arcs

This 1,910 km-long segment contains 24 volcanoes (Figures 7g,h), giving an average $d_n$ of 1.0 volcanoes/100 km (median 1.0 volcanoes/100 km, Table 2). The northern Tonga Arc section has a much higher volcano density than the southern Kermadec Arc section (Figures 7g,h). Volcano volumes are relatively even between 20 and 400 km$^3$, with mean and median averages of 125 km$^3$ and 112 km$^3$, respectively. $d_V$ ranges between 0 and 860 km$^3$/100 km, with an average of 126 km$^3$/100 km and a much lower median value of 31 km$^3$/100 km (Figure 7 h).

Only one feature, the Louisville hotspot, is subducting under these arcs (Koppers et al., 2004, 2011; Ito and van Keken, 2007; King and Adam 2014) (Figures 7g, h). The oldest volcanism occurred at the Osborn seamount at 76.7±0.8 Ma, near the trench where the Pacific Plate is subducting (Koppers et al., 2004, 2011). Its projection crosses the segment in between the Tonga Arc to the north and the Kermadec Arc to the south, and its swath does not include any volcano (Figures 7g, h). It is hence associated with a gap in $d_n$ and $d_V$ (Figure 7h), which is further discussed in section 4.1.

### 4. Discussion

In this section we discuss the results presented above. In particular, we discuss the increase or decrease of arc volcanism in the projected swaths of the subducting linear features. Although previous studies have used the volcano density number $d_n$ as a criterion to assess arc volcanism increase or decrease (e.g., Rosenbaum and Mo, 2011), we consider that the volcano volume density $d_V$ is a more pertinent criterion. Indeed, fewer volcanoes with large volumes may still be indicative of a higher amount of melt above the slab. Table 2 summarizes the statistics on $d_n$ and $d_V$ (mean and median values) for each subducting feature and also for each arc segment to which it belongs. To avoid the influence of outliers, in the following we consider the median value of $d_V$ to quantify variations of arc volcanism.

The comparison between the median $d_V$ within the projected swath of a linear subducting feature with the overall arc segment median value can be considered as an indicator of volcanism increase or decrease associated with the feature. The *y*-axis in Figure 8 is the ratio between $d_V$ of the feature and $d_V$ of the volcanic arc. However, this quantification is *regional*, as it considers the whole arc segment as the reference. The presence of another subducting feature with a large $d_V$ in the same arc segment could mask a possible local increase in volcanism. To avoid such a bias, we also provide a *local* description of $d_V$ variations along the volcano road. We distinguish four different cases (*x*-axis, Figure 8): (i) gap (no volcano inside the projected swath) or local minimum in $d_V$, (ii) local maximum in $d_V$, (iii) isolated volcano within the swath, and (iv) no specific pattern. This local description, including clear patterns (i-iii), bring complementary information about the effect of a subducting linear topographic feature on arc volcanism.

Furthermore, to discuss the effect of the subduction of a topographic feature on arc volcanism increase or decrease, it is mandatory to know its origin and nature. We classify the 35 subducting linear features considered in this study into eight different categories (see Table 1, and Figure 8): hotspot chains (HS), oceanic plateaus (OP), inactive or active arc ridges (iAR and aAR, respectively), inactive or active mid-oceanic ridges (iMOR and aMOR, respectively), fracture zones (FZ) and volcanic ridges (VR). Previous studies have used the generic terms "aseismic ridges" or "volcanic ridges" to describe long and linear features, composed of several volcanoes (e.g., Cross and Pilger, 1982; McGeary et al., 1985; Rosenbaum and Mo, 2011). They are generally not associated with earthquakes but are associated with thicker crust (Wilson, 1963; Morgan, 1971). However, such features are either created by the interaction of a mantle plume or a smaller-scale mantle upwelling and a moving lithosphere, and hence their origins are different.

In this discussion, we focus on each type of linear feature and the effect of its subduction on arc volcanism. Hotspot chains, for example, are created by the interaction of a mantle plume with the lithosphere drifting over it, and therefore have more enriched and evolved compositions (OIB – Ocean Island Basalt – signature). An aseismic or volcanic ridge, on the other hand, does not require the presence of a mantle plume. It can be created by passive mantle upwellings through lithospheric discontinuities or by other types of convective flows such as those created by a lithospheric step. Their geochemical signature is MORB (Mid-Oceanic Ridge Basalt)-like and differs from the hotspot signature. It is therefore expected to have a different effect on arc volcanism due to its subduction. Figure 8 summarizes the results for all subducting linear features: comparison of the median value of $d_V$ in the swath with the value of the corresponding arc segment, local pattern in $d_V$, and feature type (symbol shape and color, see legend). For clarity, the acronym of each feature has been reported above its corresponding symbol.

In the next sections (4.1 to 4.7) we discuss the results for each type of linear feature, to assess possible correlations with an increase or decrease in arc volcanism. In the last section (4.8), we integrate other independent data and models and compare with the volcano volume densities computed here.

### 4.1 Hotspot chains and Oceanic plateaus

Hotspot chains are created by the partial melting of the tail of a plume, and oceanic plateaus are created by the partial melting of a plume head (Sleep, 1992; Mann and Taira, 2004; Campbell, 2007). Both, therefore, have similar compositions that include ultramafic rocks. Oceanic plateaus are flat features, rising 2-3 km above the seafloor, and hence encompass large volumes of erupted material.

The analyzed oceanic plateaus (Ogasawara plateau, Dutton Ridge, West Torres plateau, and even the Yakutat block) are associated with arc volcanism increase (Figure 8; see Figure 5f for Yakutat). Of the 13 analyzed hotspot chains, five are associated with volcanism increase (Copiapó Ridge, Taltal Ridge, Carnegie Ridge, Cobb Seamounts, and the Emperor Seamounts), and four with volcanic gaps or local minima (Juan Fernandez Ridge, Nazca Ridge, Caroline Ridge, Louisville hotspot) (Figure 8). The remaining four hotspot chains (Iquique Ridge, Cocos Ridge, Seamount Province, Kodiak-Bowie Chain) lay in the gray areas of Figure 8, meaning that their interpretation is not straightforward.

We start by discussing these problematic features. Although Iquique Ridge does not display any specific pattern in $d_V$, it is interesting to note that its swath includes a small volcanic gap (the Pica gap). Even if it cannot be classified into the "gap / local minimum" pattern, it is consistent with the fact that $d_V$ for the feature is smaller than $d_V$ for the arc segment.

The subduction of the Cocos Ridge and the Seamount Province are associated with an isolated volcano (Barú) in their swaths and a few very large volcanoes located just outside the northwestern side of their projections (Figures 4i, j). The characteristics and the history of the Carnegie and Cocos ridges are relatively similar. They are hotspot tracks created respectively by the eastward motion of the Nazca Plate and the northeastward motion of the Cocos Plate over the Galapagos plume (Gutscher et al.,1999a). It is thus surprising that the subduction of the Carnegie Ridge is associated with a volcanism increase, whereas the subduction of the Cocos Ridge is associated with a volcanism decrease, as our study and previous works point out (e.g., Nur and Ben-Avraham, 1983; McGeary et al., 1985; Kolarksy et al., 1995). However, even if the size of the Carnegie and Cocos ridges are similar, there are several differences between these features. For example, their geochemical signature varies (Werner and Hoernle, 2003; Harpp et al., 2005). In particular, a spreading center formed along the Cocos Ridge 3.5 My ago (Harpp et al., 2005) and was only active for a short period of time (Harpp et al., 2005), but may have diluted the OIB signature. Figure 9 investigates the correlation between subducting linear features and lateral seismic velocity anomalies, *dvs*, provided by Debayle et al. (2016), at 100 km depth. The slow seismic velocities seen in Figure 9a resembles the pattern formed at other subducting MOR such as the Chile Ridge or Juan de Fuca Ridge (Figures 9b, c). The transform fault (unnamed) we follow under the Caribbean plate is the last surface expression of the diverging center between the Nazca and Cocos plates, and displays a zigzagging pattern before this last segment of transform fault. Therefore, the slow seismic anomaly seen in Figure 9a may be the signature of the Cocos-Nazca spreading center or may be the signature of the spreading center formed along the Cocos ridge 3.5 Ma ago. Alternatively, the seismically slow region could represent enhanced melting associated with the Seamount Province subduction, as proposed by Gazel et al. (2015; 2021). This last point is developed further in the discussion.

The Kodiak-Bowie Chain subduction is associated with a small local peak in $d_V$ and a large peak in $d_n$ (Figure 5f). Although the median value of $d_V$ in its swath is smaller than the arc segment's, it is of the same order of magnitude (Table 2). As discussed previously, other phenomena generating large $d_V$ in the same arc segment may mask a local volcanism increase. Hence, the local peak for Kodiak-Bowie could be interpreted as a minor increase in volcanism.

Now, we focus our discussion on the linear features associated with arc volcanism decrease. Former discussions about the influence of hotspot chain and oceanic plateau subduction considered the buoyancy of these features, generally associated with thickened crust. Because they are buoyant, they are difficult to subduct and sometimes their subduction results in collisions or lateral growth (Niu et al., 2003, 2015; Kerr, 2014; Zhang et al., 2014; Whattam and Stern, 2015). Other studies invoke slab shallowing. As the dip value decreases, the pressure-temperature conditions do not allow melting of the mantle above the slab (Kay and Mpodozis, 2002; Bishop et al., 2017). This could explain the gaps associated with the subduction of the Juan Fernandez (Figure 3d) and Nazca ridges (Figure 3j). However, even if Juan Fernandez is now associated with a volcanic gap, the locus of its subduction has migrated from north to south during the Cenozoic (e.g., Pilger 1981, 1984; Yáñez et al. 2001), and its passing below the Central Volcanic Zone of the Andes has been related to an increase in magmatism and/or southward migration or broadening of arc activity (Yáñez et al. 2001; Trumbull et al. 2006; Kay and Coira 2009; Guzmán et al. 2014; de Silva and Kay 2018). The subduction of the buoyant Nazca Ridge may be at the origin of the low-angle slab dip and therefore at the origin of the Peruvian flat slab, as suggested by previous studies (e.g., Pilger 1981; Gutscher et al., 2000; Antonijevic et al., 2015; Bishop et al. 2017). The Caroline Ridge is situated away from any volcanic arc (Figure 6f). Its convergence is thought to have resisted subduction and produced a collision (e.g., McGeary et al., 1985; van Rijsingen et al, 2019). Alternatively, and independently of the presence of the Caroline Ridge, the pressure-temperature conditions needed to produce melt may not be achieved along this subduction zone. Indeed, there are no arc volcanoes over a wide region, not only over the subduction of the Caroline Ridge. The southernmost volcano belonging to the Izu-Bonin-Mariana Arc is at latitude 13.25°N, whereas the Caroline Ridge intersects the trench at latitude 10.5°N. Therefore, there may be no connection between the subduction of the Caroline Ridge and the lack of arc volcanoes.

The subduction of the Louisville hotspot is associated with a volcanism gap. This gap has been previously indicated by Nur and Ben Avraham (1983), Ballance et al. (1989) and Rosenbaum and Mo (2011). However, there are submarine volcanoes within the swath that have been studied by Massoth et al. (2007) and Timm et al. (2013), but that are not in our database because they are not included in the Smithsonian Institute Global Volcanism Program database of Holocene volcanoes (they are included in the Smithsonian Institute's Pleistocene database; see Siebert et al. 2010). Thus, the interpretation of a gap related to the subduction of Louisville hotspot is not straightforward.

Several of the subducting hotspot chains and all the oceanic plateaus considered in this study are however associated with a clear arc volcanism increase (Figure 8). Such an increase could be related to the fact that the isotope, trace elements, and volatile compositions of hotspot chains and oceanic plateaus (i.e., OIB composition) differ from the MORB (Hofmann, 2007; Alderton and Elias, 202). Gazel et al. (2015; 2021) show that the subduction of the Galápagos hotspot track creates on the continental side, in Costa Rica, arc magmas with isotopic and trace element compositions displaying an OIB affinity, similar to the Galápagos-OIB lavas. According to the authors, the Galápagos track subduction creates a more fertile source for arc volcanoes and increases magma productivity.

OIBs also have higher water contents than MORB (Reekie et al., 2019; Alderton and Elias, 2021). However, hydrothermal alteration potentially changes that relationship, adding considerable amounts of water to the oceanic crust. Moreover, recent work suggests that the faults created by bending of the slab during subduction provide pathways for fluid flow that could extend into the sub-slab mantle (Miller et al., 2020). Even assuming that a subducting hotspot track contains more water, the chemical and physical phenomena occurring in subduction zones are complex. Volatiles influence the eruption style, and the volume of erupted magma. The outgassing starts earlier (i.e., deeper) for higher initial volatile contents (Zellmer et al. 2015), increasing the viscosity of the ascending magma (Sparks et al. 2000; Cashman 2004), and hampering its rise. Therefore, when the subducting slab has a high water content, the magma stalls at bigger depths on its way to the surface (Zellmer 2009), and produces longer eruptions (Edmonds et al. 2014) with potentially more magma erupted. However, these longer eruptions occur with a lower frequency because the magma is more compressible (Zellmer et al. 2015). Besides, the eruption dynamics is controlled by other parameters such as the magma ascent speed and permeability. Therefore, the relationship between the amount of volatile subducting and the volume of arc volcanoes is not straightforward (Zellmer et al. 2015).

The chemical signature of OIBs (i.e., a more enriched source and/or more volatiles) could explain why the subduction of all the studied oceanic plateaus and of several hotspot chains are associated with an increase in arc volcanoes volumes, $d_V$. Such a direct correlation has not been investigated yet in a systematic way, although local studies show insightful correlations (e.g., Gazel et al., 2015; 2021; Dorendorf et al., 2000; Nishizawa et al., 2017). In the future, it would be worth being further investigated, both with global and local studies, together with the correlation between geochemical and geophysical data.

We could then wonder why the subduction of all the hotspot chains are not associated with a $d_V$ increase. The OIBs signature differs from the MORB ones, in terms of isotopes and volatiles for example, but OIBs themselves have very heterogeneous signatures (Hofmann, 2007; Alderton and Elias, 2021, 2021). Even if hotspot chains have a similar origin, as they are created by the drifting of an oceanic plate over a mantle plume (Crough, 1983; Sleep, 1990), their chemical signature varies (Hofmann, 2007; Fitton et al., 2021). In addition, even with a similar origin, regional effects, such as the upper plate structure and tectonic setting may be important.

Concurrently, there may also be an interplay between chemical and thermo-mechanical effects. On the one hand, hotspot chains are buoyant and may resist subduction, sometimes resulting in collisions or lateral growth (e.g., Whattam and Stern, 2015). If the feature subducts, its thick and cold crust changes the pressure-temperature conditions in the subduction zone, and may flatten the slab, inhibiting melting, eventually creating a volcanic gap. We refer to these two effects as thermo-mechanical effects. On the other hand, hotspot chains may constitute a more fertile source that may promote melting in subduction zones (i.e., chemical effect). Many of the hotspot chains considered in this study are associated with arc volcanism increase, meaning that the chemical effects may prevail over the thermal effects in these cases. However, the comparison between $d_V$ and the thickness of the subducting crust (see section 4.8) does not show a clear pattern. Other parameters, such as the geochemical signature of the subducting material should be taken in consideration in further studies.

In summary, out of the 13 analyzed hotspot chains, five are associated with volcanism increase (Copiapó Ridge, Taltal Ridge, Carnegie Ridge, Cobb Seamounts, Emperor Seamounts). Four are related with volcanic gaps (Juan Fernandez Ridge, Nazca Ridge, Caroline Ridge, Louisville hotspot), but the subduction of the Juan Fernandez Ridge was linked with volcanism increase in the past. The Kodiak-Bowie Chain may also be associated with a minor volcanism increase. The subduction of the analyzed oceanic plateaus (Ogasawara plateau, Dutton Ridge, West Torres plateau, and Yakutat block) is associated with arc volcanism increase. We propose that these different patterns are created by the interplay between chemical (which may enhance melting) and thermo-mechanical effects (inhibiting melting), and/or by the inherent variations in the geochemical compositions of hotspot chains. Some of the volcanic gaps associated with hotspot chains (e.g., Caroline Ridge) may actually not be related to the subduction of these features. The Cocos Ridge and the Seamount Province are difficult to interpret because of the superposition of several phenomena: subduction of both hotspot chains and transform fault, and also the presence of a seismically slow region. Such an anomaly could be associated with the subduction of an active mid-ocean ridge, which may create a slab window, or to enhanced melting associated with the Seamount Province subduction.

### 4.2 Volcanic ridges

As stated previously, volcanic ridges are volcanic features whose origin is still debated. They are susceptible of having a MORB-like signature, as opposed to hotspots who have a more enriched, OIB-like signature. As discussed in section 3, sometimes the origin of the linear oceanic features is not known. In these cases, we will use the generic term "volcanic ridges", as it is more likely that its geochemical signature should be MORB-like rather than OIB. Our study includes four volcanic ridges: the Sarmiento Ridge, the Alvarado Ridge, the scarp south of Grijalva Fracture Zone, and the Morozko Seamounts. The subductions of the Alvarado Ridge, the scarp south of the Grijalva Fracture Zone, and the Morozko Seamounts are associated with small volcanism increases (Figure 8). This may be related to the strong hydration of these features, promoting melt generation. Former studies indicate that enriched slab-derived fluids are released mainly in the vicinity of subducted features (Peate et al., 1997; Baillard et al., 2018). However, the conclusion is not clear for the Alvarado Ridge and the scarp south of the Grijalva FZ, due to the subduction of the Carnegie Ridge in the same region, which may be dominant in controlling magmatic production. On the other hand, the Sarmiento Ridge is located on the northern side of the Peruvian flat slab and its subduction is not associated with any arc volcano, although there may be no connection between the lack of arc volcanoes and this linear feature.

### 4.3 Active mid-oceanic ridges

Mid-oceanic ridges are divergent margins where new seafloor is being created. Active mid-oceanic ridges are associated with thinner crust. Three out of the four analyzed active mid-oceanic ridges (Chile Ridge, Cocos-Nazca Ridge and Juan de Fuca Ridge) are associated with arc volcanism gaps (Figure 8). In particular, the subduction of the Chile Ridge is clearly associated with the Patagonian Volcanic Gap, occurring between the Southern and Austral volcanic zones of the Andes, as reported in previous studies (Orihashi et al., 2004; Stern, 2004), although its swath does include a few volcanoes on the sides. Rosenbaum and Mo (2011) find that the subduction of Woodlark Spreading Center is associated with a decrease in $d_n$. We also find that it is associated with an decrease in both $d_n$ and $d_V$, when comparing the mean values of the feature to the Solomon arc (Table 2). The median values of $d_n$ and $d_V$ are null along the Woodlark Spreading Center swath, and along the Solomon arc.

Thorkelson (1996) and Sisson et al. (2003) propose that the subduction of active mid-oceanic ridges create slab windows along the spreading axis of the mid-oceanic ridge. In addition to the corner flow, generally occurring above the subducted slab, a broad upwelling asthenospheric flow would occur in such cases.

Above slab windows, arc magmatism is typically interrupted and replaced by back-arc intraplate-type volcanism (Hole et al. 1991; Thorkelson 1996; Thorkelson et al. 2011). We have investigated the existence of slab windows along the four active mid-oceanic ridges considered in this study. In Figure 9, we report the lateral variations of the seismic velocity anomalies, dvs, provided by the tomography model developed by Debayle et al. (2016) at 100 km depth.

The subductions of the Juan de Fuca and Chile ridges are associated with negative *dvs*, which can be interpreted as higher temperatures (Figures 9b,c). For the Cocos-Nazca Ridge, we observe a similar pattern under the Cocos Ridge and the unnamed transform fault, which is the last surface expression of the subducting Cocos-Nazca Ridge (Figure 9a). We suggest this seismically slow region could be created by the Cocos-Nazca Ridge subduction. Our results support the scenario proposed by Thorkelson (1996) and Sisson et al. (2003): the subduction of active mid-oceanic ridges are associated with slab windows and gaps in arc volcanism.

Along the Woodlark Spreading Center there is no noticeable seismically slow region above the subducting slab (Figure 9d). The apparent lack of a slab window in this region could be explained by poor resolution of the tomography models, as other data suggest that the Woodlark Spreading Center subduction is indeed associated with a slab window (Chadwick et al., 2009; Schuth et al., 2011). The geochemical signature of lavas, such as trace elements and isotopic characteristics, especially their Pb composition, can bring insights on the processes occurring under the surface. In particular, such data can indicate the presence of slab windows, that allow mixing between arc mantle and N-MORB (normal MORB, depleted in incompatible elements, generally found along oceanic divergent margins) end-members, thus forming transitional lavas. Such a scenario has been invoked to account for the geochemical signature of the Woodlark Spreading Center (Chadwick et al., 2009; Schuth et al., 2011). In particular, the three volcanoes located on the subducting plate, Simbo, Kana Keoki and Coleman Seamount, have an arc volcano geochemical signature (Chadwick et al., 2009; Schuth et al., 2011). According to these authors, this signature indicates transfer of arc mantle across the subduction zone through the slab window. So even if the Woodlark Spreading Center subduction is not associated with slow seismic velocities, geochemical data and arc volcanoes densities denote the presence of a slab window.

A similar scenario has been invoked to account for the subduction of the Chile Ridge under South America (Klein and Karsten, 1995). Indeed, the basalts along the Chile Ridge have arc-like trace element and isotopic characteristics. This also indicates transfer of arc mantle across the subduction zone through a slab window. Johnston and Thorkelson (1997) explain the presence of the Galapagos plume geochemical characteristics in lavas on the Caribbean plate through a similar flow, occurring through the slab window created by the Cocos-Nazca Ridge subduction. In summary, the four considered active mid-oceanic ridges seem to be associated with slab windows.

### 4.4 Inactive mid-oceanic ridges

The Phoenix Ridge is an inactive mid-oceanic ridge. It is located at the extremity of the Austral Volcanic Zone of the Andes, 400 km east of the Fueguino volcano, and therefore not associated with volcanism. There may be no connection between its subduction and the end of the volcanic zone.

### 4.5 Active arc volcano ridges

Active arc volcano ridges are chains of arc volcanoes along which volcanism is still active. As they are associated with thicker crust of relatively low density, they are considered to be difficult to subduct, and could lead to accretion and lateral growth (Niu et al., 2015; Kerr, 2014; Whattam and Stern, 2015). However, McGeary et al. (1985) show that some active arc volcano ridges have subducted. The subduction of the Izu-Bonin-Mariana Arc under the Honshu Arc is associated with enhanced arc volcanism (Figure 8). As discussed in section 3, Ichihara et al. (2017) speculated that Mt. Fuji, the largest volcano in this region, is fed by what was an arc volcano of the Izu-Bonin-Mariana Arc. The seismic structures of this complex region have started being resolved only recently (Kinoshita et al., 2015; Kashiwagi et al., 2020), and support this scenario.

### 4.6 Inactive arc volcano ridges

Inactive arc volcano ridges are composed of arc volcanoes that are no longer associated with active volcanism because they have been isolated from active subduction through various phenomena such as back-arc extension (Hickey-Vargas, 2005; Ishizuka et al., 2018). Our study encompasses five inactive arc volcano ridges: D'Entrecasteaux, Kyushu-Palau, Amami, south Amami, and Daito ridges. The D'Entrecasteaux and Kyushu-Palau ridges are associated with arc volcanism increase (Figure 8). Although the subduction of the Kyushu-Palau ridge has been associated to a small-scale volcanic gap between Aso and Kirishima (e.g., Hata et al,. 2014), our projection links it to the northern volcanoes, including Aso. Our results are in agreement with Mahony et al. (2011), who report that the volcanism increase in the Aso region is related to the Kyushu-Palau ridge subduction. The subduction of the D'Entrecasteaux ridge has been studied by Rosenbaum and Mo (2011), who concluded it is associated with a decrease in arc magmatism. Our projection suggests that its subduction is associated with the large Aoba volcano and hence with an increase of arc volcanism (Figures 7e, f). Considering only $d_n$ would not allow to resolve such an increase, because Aoba is quite isolated (Figures 7c, e). Note that our projection of the D'Entrecasteaux ridge corresponds to the location constrained from intermediate-depth seismicity by Baillard et al. (2018).

The subductions of the Amami, south Amami, and Daito ridges are associated with a decrease in arc volcanism. More precisely, these three features are located on the southern extremity of the Kyushu-Ryuku volcanic zone. Previous studies (e.g., Okamura et al. 2017) show that the Amami, Daito and Kyushu-Palau ridges migrated north and west in the last 23 Ma. There is no evidence that the end of arc magmatism is related to the subduction of these features rather than to other larger scale phenomena. Away from volcanic gap edges, we find that inactive arc volcano ridges subductions are associated with enhanced arc volcanism. Note that the swath of Amami, south Amami and Daito ridges is quite small, and the volcanoes are rather sparse in this region. Therefore, the fact that the subduction of these features is associated with an increase or decrease in arc volcanism strongly depends if the projection includes a single volcano or not. The results concerning these three ridges may thus be taken with caution, as they may not be statistically significant.

### 4.7 Fracture zones

Oceanic fracture zones are created by the offset of spreading centers. As fracture zones are associated with topographic offsets, they promote seawater infiltration, and consequently the hydrothermal alteration of peridotites into serpentinites (Manea et al., 2014). Water-rich serpentinite will increase the volume of slab-derived fluids, thus affecting the melt under arc volcanoes (Ulmer and Trommsdorff, 1995; Kerrick, 2002; Manea et al., 2014). In theory, more melt would be at the origin of enhanced volcanism (Cooper et al., 2020), although previous studies have not directly demonstrated such correlation.

Four fracture zones have been considered in this study: the Mendana and the Grijalva Fracture Zones, the transform fault subducting under the Caribbean plate, and the Tehuantepec Ridge. The subduction of the Mendana Fracture Zone is associated with a volcanic gap (the Peruvian flat-slab). It is however unlikely that the subduction of this narrow fracture zone would be at the origin of the slab flattening, which has rather been associated with the subduction of the Nazca Ridge (see above). Hence, we cannot discuss the effect of the Mendana Fracture Zone subduction on arc volcanism, as this latter seems to be controlled by other larger scale factors.

The subduction of the Grijalva Fracture Zone is associated with a peak in volcanism. However, several subducting features overlap in this region (Northern Volcanic Zone of the Andes) (Figures 4a, g), in particular the wide Carnegie Ridge, and it is difficult to isolate the signature of each feature. The transform fault subducting under the Caribbean plate is associated with a decrease in arc volcanism (Figure 8) but does not display any local pattern in $d_V$. Here again, several features overlap in a relatively small area (Central America Arc) (Figures 4i, j, l). The subduction of the Tehuantepec Ridge correlates with an isolated small volcano, El Chichón.

Although it has been shown that the subduction of fracture zones affects the geochemical signature of associated volcanoes (e.g., Manea et al., 2014), no clear pattern is found between fracture zone subduction and arc volcanism output. This is probably because only a few fracture zones have been integrated in this study. As stated in the introduction, our features selection was based on bathymetry data. It is generally difficult to pick fracture zones from bathymetry only. Wessel et al. (2015) designed a fracture zone database by integrating seafloor fabric data, magnetic lineations, and plate tectonic models. They find almost sixty fracture zones intersecting the subduction zones in the Pacific basin. As bathymetry data only allow the thorough characterization of four fracture zones, we cannot derive a statistic study on the effect of fracture zones subduction. Future studies should investigate this effect more systematically, by taking into account other data and databases.

### 4.8 Comparison with other data

Figure 10 shows the comparison between $d_V$ and other data. As two outliers dominate, in the SOM we also provide this figure without these two features (Figure S1), to facilitate the visual comparison and investigate the possible correlations. The comparison between $d_V$ and $d_n$ is showed in Figure 10a. These variables are not correlated. For example, the D'Entrecasteaux ridge and the West Torres Plateau have large $d_V$ but relatively small $d_n$, as their subduction is associated with a few large volcanoes. As discussed above, we consider that $d_V$ is a better criterion to discuss arc volcanism enhancement or decrease, as it is a better proxy of the volume of melt.

We also present correlations between $d_V$ and other parameters, often invoked in discussions about arc volcanism variations (Figures 10b-h). These parameters are: α, the angle between the linear oceanic feature and the perpendicular to the trench (Figure 10b); δ, the slab dip (Figure 10c), computed from Slab2 (Hayes et al., 2018); the slab age (Figure 10d, data from Hughes and Mahood, 2011); the difference between the kinematic velocity of the subducting plate and the kinematic velocity of the overriding plate, $v_{kin}$ (calculated with the UNAVCO Plate Motion Calculator using GRSM 2.1 model, Kreemer et al., 2014) from which we derive the norm of this vector: $|v_{kin}|$ (Figure 10e); and the crustal thickness along the subducting feature, $Tc_{oceanic}$, (Figure 10f, data from Pasyanos et al., 2014).

We have also considered the effect of the upper plate structure on magma productivity. Figure 10g shows comparison between $d_V$ and the tectonic regimes in the upper plate, and Figure 10h plots $d_V$ and the crustal thickness of the upper plate. These data are extracted from Hughes and Mahood (2011), who gathered data from literature review and datasets to obtain several parameters describing arc segments. Previous studies (e.g., Karlstrom et al., 2014) show that arc magmatism is related to parameters such as the crustal thickness of the upper plate, $Tc$. However, we do not find any clear correlation between $d_V$ and any of these parameters (Figure 10). We find a tentative correlation between *dv* and slab age: all the linear features for which we found a decrease in arc volcanism are associated with a slab age younger than 54 Ma (with the exception of Louisville hotspot). However, not all the features associated with a slab younger than 54 Ma are characterized with a decrease in arc volcanism. Variations of arc volcano volumes are probably controlled by the interplay of several parameters.

## 5. Conclusions

We have studied the correlation between subducting linear oceanic features and arc volcanism around the Pacific basin. Our analysis considers the inland projections of the subducting features, taking into account the subduction zone geometry and integrating the latest data and models, such as an updated arc volcano database and the latest slab dip angles (provided by Slab2.0, Hayes et al., 2018). The original aspects of our work are the computation of a continuous volcano road, formed by arc volcanoes, and the quantification of both the volcanoes number and volume densities along this volcano road. These regional statistics are also compared to the local patterns observed in volume densities to thoroughly assess variations in arc volcanism.

The subduction of oceanic plateaus (Ogasawara Plateau, Dutton Ridge, West Torres Plateau) is associated with a volcanism increase. Out of the thirteen analyzed hotspot chains, five are clearly associated with a volcanic increase (Copiapó Ridge, Taltal Ridge, Carnegie Ridge, Cobb Seamounts, Emperor Seamounts), one was in the past (Juan Fernandez Ridge) and one displays a local maximum in the volcano volume density (Kodiak-Bowie). On the other hand, three are related with volcanic gaps (Nazca Ridge, Caroline Ridge, and the Louisville hotspot). We propose that these different patterns reflect the interplay between chemical effects (enhancing melting) and thermo-mechanical effects (inhibiting melting), and/or by inherent variations in the geochemical composition of hotspot chains. The subduction of volcanic ridges is generally associated with a small increase in arc volcanism (for three out of the four considered volcanic ridges), which may be accounted for by the fact that these features are highly hydrated and therefore promote melt. The four considered active mid-oceanic ridges (Chile Ridge, Cocos-Nazca, and Juan de Fuca ridges, and the Woodlark Spreading Center) are associated with slab windows, although their signature in geophysical data is clearly supporting this scenario for only three of them (Chile Ridge, Cocos-Nazca, and Juan de Fuca ridges). The only analyzed inactive mid-oceanic ridge (the Phoenix Ridge) is located in a volcanic gap, at the extremity of an arc volcano segment, whereas the only analyzed active arc volcano ridge (the Izu-Bonin-Mariana Arc) is associated with arc volcanism increase. Out of the five inactive arc volcano ridges considered, three (Amami, south Amami, and Daito ridges) are located near a volcanic arc extremity and associated with volcanism decrease, without any clear causality. Away from volcanic arc edges, we find that the subduction of inactive arc volcano ridges is associated with enhanced arc volcanism (D'Entrecasteaux and Kyushu-Palau ridges). No clear pattern seems to arise when studying the effect of fracture zone subduction on arc volcanism volumes, but this may be because only four fracture zone can be characterized from bathymetry data. The comparison between arc volcanism variations and other independent geophysical data (e.g., crustal thickness, stresses in the upper plate, kinematic velocities) do not exhibit any systematic correlations.

In further studies, it would be worth comparing how our projected subducted features are correlated with seismicity (if such seismic studies are available). It would also be interesting to investigate the possible correlations between arc volcano volumes and variations in the geochemical signature of arc lavas along the volcano road. This would help inferring or differing the hypotheses discussed here and in previous studies.

**Acknowledgments:**
The original manuscript benefited from the revisions of two anonymous reviewers and the editorial handling of Marie Edmonds. Datasets obtained from this research are included in this paper.

**Figures captions**

**Figure 1:** Bathymetry map (data from Tozer et al., 2019) showing the 35 subducting linear features considered in this study (solid black lines) and their projection after being subducted (dotted lines) (a) Pacific basin; (b-d) zoom-ins to the regions delimited by white rectangles in (a). The acronyms of each linear features (see Table 1) are reported next to each feature: PhR (Phoenix Ridge), ChR (Chile Ridge), JF (Juan Fernandez Ridge), CopR (Copiapó Ridge), TaR (Taltal Ridge), IqR (Iquique Ridge), NaR (Nazca Ridge), MeFZ (Mendana Fracture Zone), SaR (Sarmiento Ridge), AlR (Alvarado Ridge), sGrFZ (scarp south of Grijalva Fracture Zone), GrFZ (Grijalva Fracture Zone), CaR (Carnegie Ridge), CoNaR (Cocos-Nazca Ridge), CoR (Cocos Ridge), TF (transform fault, unnamed), SmPr (Seamount Province), TeR (Tehuantepec Ridge), JFR (Juan de Fuca Ridge), KB (Kodiak-Bowie Chain), CoSm (Cobb Seamounts), EmSm (Emperor Seamounts), MoSm (Morozko Seamounts), IBM (Izu-Bonin Mariana Arc), OgP (Ogasawara Plateau), DuR (Dutton Ridge), CaroR (Caroline Ridge), KP (Kyushu Palau Ridge), Am (Amami Plateau), sAm (South Amami Plateau), DaR (Daito Ridge), WSC (Woodlark Spreading Center), WTP (West Torres Plateau), DER (D'Entrecasteaux Ridge), Lou (Louisville hotspot). Arc volcanoes are reported by red triangles and plate boundaries, taken from Bird (2003), are reported in gray.

**Figure 2:** (a) Sketch of a subducting linear feature (thick gray line) that intersects the trench with an angle α from the perpendicular to the trench (black dashed line). δ is the slab dip angle. (b) 2D projection on the longitude/latitude map of the subducting linear oceanic feature. The subducted part of the linear feature, located under the overriding plate, is at an angle β from the perpendicular to the trench. (c) Illustration of the arc volcanoes (red triangles) and of the volcano road defined in this study (black line). A disk (here in blue) of diameter $d$=100 km (the blue circle) is translated along the volcano road in order to compute the volcanoes number and volume densities ($d_n$ and $d_V$, see text).

**Figure 3:** (a-e) Austral and Southern volcanic zones of the Andes, and (f-k) Central volcanic zone of the Andes. (a, f) General maps showing the location of arc volcanoes and of the linear oceanic features (dashed black lines) and their projected swaths (solid lines); bathymetry data from Tozer et al. (2019). (b-d) and (g-i) zoom-ins showing the location of arc volcanoes, the volcano road (gray path) and the projection of the subducting linear feature swaths (black and color lines). The color scale represents the volcano volumes (in logarithmic scale). Prominent volcanoes and other volcanoes discussed in the text are labeled. (e, k) Plots of the volcano number density ($d_n$) and volcano volume density ($d_V$) as a function of latitude; the gray

boxes represent the projections of the subducting linear feature swaths under the volcano road; the gray scale represents the number of overlapping subducting features. PhR: Phoenix Ridge, ChR: Chile Ridge, JF: Juan Fernandez Ridge, CopR: Copiapó Ridge, TaR: Taltal Ridge, IqR: Iquique Ridge, NaR: Nazca Ridge.

**Figure 4:** Same as Figure 3 for the Northern Volcanic Zone of the Andes (a-g), the region between the Northern zone and Central America (h), and Central America (i-l). SaR: Sarmiento Ridge, AlR: Alvarado Ridge, sGrFZ: scarp south of Grijalva Fracture Zone, GrFZ: Grijalva Fracture Zone, CaR: Carnegie Ridge, CoNaR: Cocos-Nazca Ridge, CoR: Cocos Ridge, TF: transform fault (unnamed), SmPr: Seamount Province, TeR: Tehuantepec Ridge.

**Figure 5:** Same as Figure 3 for the Cascades and Wrangell volcanic arcs (a-b), the Aleutian-Alaska and Wrangell volcanic arcs (c-f), and the Kamchatka-Kuril-Hokkaido volcanic arc (g-j). JFR: Juan de Fuca Ridge, KB: Kodiak-Bowie Chain, CoSm: Cobb Seamounts, EmSm: Emperor Seamounts, MoSm: Morozko Seamounts.

**Figure 6:** Same as Figure 3 for the Honshu volcanic arc (a-b), the Izu-Bonin-Mariana arcs (c-g), and the Kyushu-Ryukyu volcanic arc (h-m). IBM: Izu-Bonin Mariana Arc, OgP: Ogasawara Plateau, DuR: Dutton Ridge, CaroR: Caroline Ridge, KP: Kyushu-Palau Ridge, Am: Amami Plateau, sAm: South Amami Plateau, DaR: Daito Ridge.

**Figure 7:** Same as Figure 3 for the Solomon volcanic arc (a-b), the Vanuatu arc (c-f), and the Tonga-Kermadec arcs (g-h). WSC: Woodlark Spreading Center, WTP: West Torres Plateau, DER: D'Entrecasteaux Ridge, Lou: Louisville hotspot.

**Figure 8:** Comparison between the regional statistics (ratio between the median value of $d_V$ for each feature and the arc segment median value, see Table 2) and the local patterns in $d_V$ for each subducting linear feature (see Figure 1 and Table 2 for features acronyms). The acronyms of each linear features (see Table 1) are reported next to each feature: PhR (Phoenix Ridge), ChR (Chile Ridge), JF (Juan Fernandez Ridge), CopR (Copiapó Ridge), TaR (Taltal Ridge), IqR (Iquique Ridge), NaR (Nazca Ridge), MeFZ (Mendana Fracture Zone), SaR (Sarmiento Ridge), AlR (Alvarado Ridge), sGrFZ (scarp south of Grijalva Fracture Zone), GrFZ (Grijalva Fracture Zone), CaR (Carnegie Ridge), CoNaR (Cocos-Nazca Ridge), CoR (Cocos Ridge), TF (transform fault, unnamed), SmPr (Seamount Province), TeR (Tehuantepec Ridge), JFR (Juan de Fuca Ridge), KB (Kodiak-Bowie Chain), CoSm (Cobb Seamounts), EmSm (Emperor Seamounts), MoSm (Morozko Seamounts), IBM (Izu-Bonin Mariana Arc), OgP (Ogasawara Plateau), DuR (Dutton Ridge), CaroR (Caroline Ridge), KP (Kyushu Palau Ridge), Am (Amami Plateau), sAm (South Amami Plateau), DaR (Daito Ridge), WSC (Woodlark Spreading Center), WTP (West Torres Plateau), DER (D'Entrecasteaux Ridge), Lou (Louisville hotspot). The symbol shape and color indicates the feature origin (see legend): hotspots (HS), oceanic plateaus (OP), volcanic ridges (VR), active mid-oceanic ridges (aMOR), inactive mid-oceanic ridges (iMOR), active arc ridges (aAR), inactive arc ridges (iAR), and fracture zones or transform faults (FZ). The correlation between the regional and local information makes possible to define robust regions of arc volcanism increase (red zone) or decrease (blue zone). In the gray regions, the interpretation is not straightforward (see text).

**Figure 9:** Correlation between the projection of subducting linear features and seismic velocity anomalies at 100 km depth (from Debayle et al. (2016)). (a) Northern Andes and Central America; (b) Cascades Arc; (c) Southern Andes; (d) Solomon Arc.

**Figure 10:** Comparison between $d_V$ and $d_n$ (a), and between $d_V$ and other geophysical data (b-f): (b) α the angle between the linear feature and the perpendicular to the trench; (c) δ the slab dip), computed from Slab2 (Hayes et al., 2018); (d) the slab age (data from Hughes and Mahood, 2011); (e) the norm of the difference between the kinematic velocity of the subducting plate and the kinematic velocity of the overriding plate, |v$_{kin}$| (calculated with the UNAVCO Plate Motion Calculator using GRSM 2.1 model, Kreemer et al., 2014); (f) the crustal thickness along the subducting feature, $Tc_{oceanic}$ (data from Pasyanos et al., 2014); (g) the tectonic regimes in the upper plate: compression (-3 to -1, magnitude connotes intensity), neutral or transverse (0), tension (1 to 2) or backarc spreading (3) (data from Hugues and Mahood, 2015); (h) crustal thickness of the upper plate (data from Hugues and Mahood, 2015). The symbol shape and color indicates the feature origin (see legend): hotspots (HS), oceanic plateaus (OP), volcanic ridges (VR), active mid-oceanic ridges (aMOR), inactive mid-oceanic ridges (iMOR), active arc ridges (aAR), inactive arc ridges (iAR), and fracture zones or transform faults (FZ).

**Tables captions**

**Table 1:** Feature name, acronym, type, subducting plate, upper plate, coordinates ($X_t$, $Y_t$) of the intersection between the linear oceanic feature and the trench, width *w* of the subducting feature, dip angle δ of the subducting plate, angle τ$_E$ of the trench orientation (relative to the East direction), angle α between the perpendicular to the trench and the linear oceanic feature (Figures 2a,b), and angle β between the projection of the subducted linear feature at the surface and the perpendicular to the trench (Figures 2a,b). The gray lines with bold text indicate the arc segments to which the linear subducting features belong. The type of each feature reads as follows: hotspot (HS), oceanic plateau (OP), inactive arc ridge (iAR), active arc ridge (aAR), inactive mid-oceanic ridge (iMOR), active mid-oceanic ridge (aMOR), fracture zone (FZ), volcanic ridge (VR).

**Table 2:** Feature name, acronym, type, mean and median values of $d_n$ and $d_V$ for each arc segment (gray lines with bold characters) and each subducting linear feature.

**SUPPLEMENTAL MATERIAL**

**Figure S1:** Same as Figure 10 but without the two outliers WTP and DER.

**Table S1:** Database containing the geographical locations and volume estimates of the analyzed arc volcanoes around the Pacific basin.

**Table S2:** Feature name, acronym, type, subducting plate, upper plate, east ($v_x$) and north ($v_y$) components of the difference between the kinematic velocity of the subducting plate and the kinematic velocity of the overriding plate, |v$_{kin}$| (computed with the UNAVCO Plate Motion Calculator using GRSM 2.1 model,

Kreemer et al., 2014); slab age (data from Hughes and Mahood, 2011); crustal thickness along the subducting feature, $T_c^{oc}$ (data from Pasyanos et al., 2014); crustal thickness of the upper plate, $T_c^{cont}$ (data from Hugues and Mahood, 2015); stress code indicating the tectonic regime in the upper plate: compression (-3 to -1, magnitude connotes intensity), neutral or transverse (0), tension (1 to 2) or backarc spreading (3) (data from Hugues and Mahood, 2015). The type of each feature reads as follows: hotspot (HS), oceanic plateau (OP), inactive arc ridge (iAR), active arc ridge (aAR), inactive mid-oceanic ridge (iMOR), active mid-oceanic ridge (aMOR), fracture zone (FZ), volcanic ridge (VR).